\documentclass[aps,11pt,showpacs,a4paper]{revtex4}

\usepackage{amsfonts}
\usepackage{amsmath}
\usepackage{amssymb}
\usepackage{graphicx}

\begin{document}

\title[The moving boundary problem in the presence of a dipole magnetic field%
]{The moving boundary problem in the presence of a dipole magnetic field}
\author{H. B. Nersisyan}
\email{hrachya@irphe.am}
\author{D. A. Osipyan}
\affiliation{Division of Theoretical Physics, Institute of Radiophysics and Electronics,
National Academy of Sciences of Armenia, 378410 Ashtarak, Armenia}
\pacs{03.50.De, 41.20.Gz, 41.20.Jb, 52.30.-q}

\begin{abstract}
An exact analytic solution is obtained for a uniformly expanding, neutral,
infinitely conducting plasma sphere in an external dipole magnetic field.
The electrodynamical aspects related to the radiation and transformation of
energy were considered as well. The results obtained can be used in
analyzing the recent experimental and simulation data.
\end{abstract}

\maketitle

\section{Introduction}

\label{Intro} Many processes in physics involve boundary surfaces which
requires the solution of boundary and initial value problems. The
introduction of a moving boundary into the physics usually precludes the
achievement of an exact analytic solution of the problem and recourse to
approximation methods is required \cite{rog89,mor53} (see also \cite{jac75}
and references therein). In the case of a moving plane boundary a
time-dependent translation of the embedding space immobilizes the boundary
at the expense of the increased complexity of the differential equation. It
is the aim of this work to present an example of a soluble moving boundary
and initial value problem in spherical geometry.

The problems with moving boundary arise in many area of physics. One
important example is sudden expansion of hot plasma with a sharp boundary in
an external magnetic field which is particularly of interest for many
astrophysical and laboratory applications (see, e.g., \cite{zak03} and
references therein). Such kind of processes arise during the dynamics of
solar flares and flow of the solar wind around the earth's magnetosphere, in
active experiments with plasma clouds in space, and in the course of
interpreting a number of astrophysical observations [3-9]. Researches on
this problem are of considerable interest in connection with experiments on
controlled thermonuclear fusion \cite{sgr76} (a recent review \cite{zak03}
summarizes research in this area over the past four decades).

To study the radial dynamics and evolution of the initially spherical plasma
cloud both analytical and numerical approaches were developed (see, e.g.,
[3-9] and references therein). The plasma cloud is shielded from the
penetration of the external magnetic field by means of surface currents
circulating inside the thin layer on the plasma boundary. Ponderomotive
forces resulting from interaction of these currents with the magnetic field
would act on the plasma surface as if there were magnetic pressure applied
from outside. After some period of accelerated motion, plasma gets
decelerated as a result of this external magnetic pressure acting inward.
The plasma has been considered as a highly conducting matter with zero
magnetic field inside. From the point of view of electrodynamics it is
similar to the expansion of a superconducting sphere in a magnetic field. An
exact analytic solution for a uniformly expanding, superconducting plasma
sphere in an external uniform and constant magnetic field has been obtained
in \cite{kat61}. The nonrelativistic limit of this theory has been used by
Raizer \cite{rai63} to analyze the energy balance (energy radiation and
transformation) during the plasma expansion. The similar problem has been
considered in \cite{dit00} for a plasma layer. In the present paper we study
the uniform expansion of the superconducting plasma sphere in the presence
of a dipole magnetic field. For this geometry we found an exact analytical
solution which can be used in analyzing the recent experimental and
simulation data (see \cite{win05} and references therein).

\section{Magnetostatic treatment}

%\label{sec:}

In this section we first consider the simpler example of a non-relativistic
expansion of the plasma sphere ($v\ll c$, where $v$ is the radial velocity
of the sphere) in the presence of a dipole magnetic field. Consider the
magnetic dipole $\mathbf{p}$\ and a superconducting sphere with radius $R$
located at the origin of the coordinate system. The dipole is placed in the
position $\mathbf{r}_{0}$ from the center of sphere ($R<r_{0}$). The
orientation of the dipole is given by the angle $\theta_{p}$ between the
vectors $\mathbf{p}$ and $\mathbf{r}_{0}$. Here it is convenient to
introduce the scalar magnetic potential $\psi_{0}\left( \mathbf{r}\right) $
of the dipole magnetic field which is given by

\begin{equation}
\psi_{0}\left( \mathbf{r}\right) =\frac{\mathbf{p}\cdot\left( \mathbf{r}-%
\mathbf{r}_{0}\right) }{\left\vert \mathbf{r}-\mathbf{r}_{0}\right\vert ^{3}}
.   \label{1}
\end{equation}
The dipole magnetic field is then calculated as $\mathbf{H}_{0}\left( 
\mathbf{r}\right) =-\mathbf{\nabla}\psi_{0}\left( \mathbf{r}\right) $,

\begin{equation}
\mathbf{H}_{0}\left( \mathbf{r}\right) =\frac{1}{\left\vert \mathbf{r}-%
\mathbf{r}_{0}\right\vert ^{3}}\left[ \frac{3\left( \mathbf{r}-\mathbf{r}%
_{0}\right) \left[ \mathbf{p}\cdot\left( \mathbf{r}-\mathbf{r}_{0}\right) %
\right] }{\left\vert \mathbf{r}-\mathbf{r}_{0}\right\vert ^{2}}-\mathbf{p}%
\right] .   \label{2}
\end{equation}

When the superconducting sphere is introduced into a background magnetic
field the plasma expands and excludes the background magnetic field to form
a magnetic cavity. The magnetic energy of the dipole in the excluded volume,
i.e., in the volume of the superconducting sphere is calculated as

\begin{align}
Q_{R} & =\int_{r\leqslant R}\frac{H_{0}^{2}\left( \mathbf{r}\right) }{8\pi}d%
\mathbf{r}  \label{3} \\
& =\frac{p^{2}}{32r_{0}^{3}}\left\{ \frac{\xi\left( 1-\xi^{4}\right) \left(
3\cos^{2}\theta_{p}-1\right) +8\xi^{3}\left( 1+\cos^{2}\theta _{p}\right) }{%
\left( 1-\xi^{2}\right) ^{3}}-\frac{3\cos^{2}\theta_{p}-1}{2}\ln\frac{1+\xi}{%
1-\xi}\right\} ,  \notag
\end{align}
where $\xi=R/r_{0}<1$. This energy increases with decreasing $\theta_{p}$
and reach its maximum value at $\theta_{p}=0$ or $\theta_{p}=\pi$ that is
the magnetic moment $\mathbf{p}$ is parallel or antiparallel to the symmetry
axis $\mathbf{r}_{0}$. In addition the magnetic energy $Q_{R}$ decays
rapidly with the distance $r_{0}$ and for large $r_{0}\gg R$ is given by

\begin{equation}
Q_{R}=\frac{p^{2}R^{3}}{6r_{0}^{6}}\left( 3\cos^{2}\theta_{p}+1\right) . 
\label{4}
\end{equation}
In the case when the dipole approaches to the surface of the sphere $%
r_{0}\simeq R$ the magnetic field of the dipole becomes very large and tends
to the infinity as

\begin{equation}
Q_{R}=\frac{p^{2}}{32r_{0}^{3}}\frac{1+\cos^{2}\theta_{p}}{\left(
1-\xi\right) ^{3}}.   \label{5}
\end{equation}

We turn now to solve the boundary problem and calculate the induced magnetic
field which arises near surface of the sphere due to the dipole magnetic
field. Since the sphere is superconducting the magnetic field vanishes
inside the sphere. In addition the normal component of the field $H_{r}$
vanishes on the surface of the sphere. To solve the boundary problem we
introduce the spherical coordinate system with the $z$-axis along the vector 
$\mathbf{r}_{0}$ and the azimuthal angle $\phi$ is counted from the plane ($%
xz$-plane) containing the vectors $\mathbf{r}_{0}$ and $\mathbf{p}$. Hence,
using the expressions (\ref{A2})-(\ref{A4}) the scalar potential (\ref{1})
at $r<r_{0}$ can alternatively be represented by the sum of Legendre
polynomials (see the Appendix~A for details)

\begin{equation}
\psi_{0}\left( \mathbf{r}\right) =\frac{p}{r_{0}^{2}}\left[ \sin\theta
_{p}\cos\phi\sum_{l=1}^{\infty}\left( \frac{r}{r_{0}}\right)
^{l}P_{l}^{1}\left( \cos\theta\right)
-\cos\theta_{p}\sum_{l=0}^{\infty}\left( l+1\right) \left( \frac{r}{r_{0}}%
\right) ^{l}P_{l}\left( \cos \theta\right) \right] .   \label{6}
\end{equation}
The total magnetic field which is a sum of $\mathbf{H}_{0}\left( \mathbf{r}%
\right) $ and the induced magnetic field is obtained from equation $%
\nabla\cdot\mathbf{H}=0$. Introducing the scalar potential, $\mathbf{H}%
\left( \mathbf{r}\right) =-\mathbf{\nabla}\psi\left( \mathbf{r}\right) $,
the last equation becomes $\nabla^{2}\psi\left( \mathbf{r}\right) =0$, i.e., 
$\psi\left( \mathbf{r}\right) $ satisfies the Laplace equation. We must
solve this equation with $\mathbf{H}=0$ at $r<R$\ and boundary condition

\begin{equation}
\left. H_{r}\right\vert _{r=R}=-\left. \frac{\partial\psi}{\partial r}%
\right\vert _{r=R}=0.   \label{7}
\end{equation}

We look for the solution of the Laplace equation which in a spherical
coordinate system and at $r\geqslant R$ can be written as

\begin{equation}
\psi\left( \mathbf{r}\right) =\psi_{0}\left( \mathbf{r}\right) +\frac {p}{%
r_{0}^{2}}\left[ \sum_{l=0}^{\infty}\alpha_{l}\left( \frac{R}{r}\right)
^{l+1}P_{l}\left( \cos\theta\right) +\cos\phi\sum_{l=1}^{\infty}\beta
_{l}\left( \frac{R}{r}\right) ^{l+1}P_{l}^{1}\left( \cos\theta\right) \right]
,   \label{8}
\end{equation}
where $\alpha_{l}$ and $\beta_{l}$\ are the arbitrary constants and should
be obtained from the boundary condition (\ref{7}). The second term in
equation~(\ref{8}) is the induced magnetic field. From equations~(\ref{6})-(%
\ref{8}) one finds

\begin{equation}
\alpha_{l}=-l\left( \frac{R}{r_{0}}\right) ^{l}\cos\theta_{p},\qquad
\beta_{l}=\frac{l}{l+1}\left( \frac{R}{r_{0}}\right) ^{l}\sin\theta_{p}. 
\label{9}
\end{equation}

Substituting equation~(\ref{9}) into equation~(\ref{8}) and using the
summation formula obtained in Appendix~A from (\ref{8}) we find

\begin{equation}
\psi\left( \mathbf{r}\right) =\frac{\mathbf{p}\cdot\mathbf{R}_{0}}{R_{0}^{3}}%
+\frac{\mathbf{Q}\cdot\mathbf{R}_{\ast}}{R_{\ast}^{3}}+\psi_{\mathrm{QD}%
}\left( \mathbf{r}\right) ,   \label{10}
\end{equation}
where

\begin{equation}
\psi_{\mathrm{QD}}\left( \mathbf{r}\right) =-\xi^{3}\frac{\left( \mathbf{p}%
_{\bot}\cdot\mathbf{R}_{\ast}\right) }{R_{\ast}^{3}}\left( \frac{R_{\ast}^{2}%
}{\mathbf{r}\cdot\mathbf{R}_{\ast}+rR_{\ast}}-\frac{1}{2}\right) . 
\label{11}
\end{equation}
Here $\mathbf{r}_{\ast}=\xi^{2}\mathbf{r}_{0}$, $\mathbf{R}_{0}=\mathbf{r}-%
\mathbf{r}_{0}$, $\mathbf{R}_{\ast}=\mathbf{r}-\mathbf{r}_{\ast}$,

\begin{equation}
\mathbf{p}_{\bot}=\mathbf{p}-\frac{\left( \mathbf{p}\cdot\mathbf{r}%
_{0}\right) \mathbf{r}_{0}}{r_{0}^{2}},\qquad\mathbf{Q}=\frac{\xi^{3}}{2}%
\left[ \mathbf{p}-\frac{3\left( \mathbf{p}\cdot\mathbf{r}_{0}\right) \mathbf{%
r}_{0}}{r_{0}^{2}}\right] .   \label{12}
\end{equation}
The term $\psi_{\mathrm{QD}}\left( \mathbf{r}\right) $ in equation~(\ref{10}%
) can be interpreted as a magnetic field of point-like quadrupole with the
"quadrupole moment" $D_{\alpha\beta}\left( \mathbf{r}\right) $\ and located
in the $xz$-plane inside the sphere at the distance $\mathbf{r}_{\ast}$ ($%
r_{\ast}=\xi R<R$) from the centre. At large distances this term behaves as $%
\psi_{\mathrm{QD}}\left( \mathbf{r}\right) \simeq xzD_{xz}/r^{5}$ with the
quadrupole moment $D_{xz}=\frac{r_{0}}{2}\xi^{5}p\sin\theta_{p}$ ($%
D_{\alpha\alpha}=D_{xy}=D_{yz}=0$ and $\alpha=x,y,z$). The induced electric
field is calculated from the Maxwell's equation $\nabla\times\mathbf{E}=-%
\frac{1}{c}\frac{\partial\mathbf{H}}{\partial t}$. However, if plasma radial
velocity is small, $v/c\ll1$, the amplitude of electric field is small as
well (of the order of $\frac{v}{c}H_{0}\left( \mathbf{r}\right) $) and may
be completely ignored. Below we consider two particular cases for the
magnetic dipole orientation in the space.

(i)~The case $\theta_{p}=0;~\pi$. In this case the magnetic dipole is
parallel or antiparallel to the vector $\mathbf{r}_{0}$. Obviously due to
the symmetry reason the magnetic field does not depend on $\phi$ and $%
H_{\phi}=0$. The magnetic field component $H_{\theta}=-\left( 1/r\right)
\left( \partial \psi/\partial\theta\right) $ induces the surface current on
the sphere. The ponderomotive forces resulting from interaction of this
current with the magnetic field acts on the sphere surface with a magnetic
pressure which can be calculated as an energy density of the magnetic field

\begin{equation}
P_{\parallel}\left( \theta\right) =\left. \frac{H_{\theta}^{2}}{8\pi }%
\right\vert _{r=R}=\frac{9p^{2}}{8\pi r_{0}^{6}}\frac{\left( 1-\xi
^{2}\right) ^{2}\sin^{2}\theta}{\left( \xi^{2}+1-2\xi\cos\theta\right) ^{5}}%
.   \label{13}
\end{equation}

This pressure vanishes at $\theta=0$, $\pi$ and has its maximum at

\begin{equation}
\cos\theta_{\max}=\frac{10\xi}{\sqrt{\left( \xi^{2}+1\right) ^{2}+60\xi^{2}}%
+\xi^{2}+1}.   \label{14}
\end{equation}

The value of $\theta_{\max}$ tends to zero when the dipole comes close to
the sphere and shifts toward the larger values, $\theta_{\max}\simeq\pi/2 $,
when the dipole goes to the infinity. Therefore the layer near $\theta\simeq
\theta_{\max}$ of the expanding sphere will be mainly deformed by the
external magnetic pressure. This behavior is clearly seen in the
particle-in-cell simulation \cite{ner06}.

The total force is calculated as a surface integral of the magnetic pressure

\begin{equation}
\mathcal{F}_{\parallel}=2\pi R^{2}\int_{0}^{\pi}P_{\parallel}\left(
\theta\right) \sin\theta d\theta=\frac{3p^{2}}{r_{0}^{4}}\frac{\xi^{2}\left(
1+\xi^{2}\right) }{\left( 1-\xi^{2}\right) ^{4}}.   \label{15}
\end{equation}
This force behaves as $\mathcal{F}_{\parallel}\sim l^{-s}$ with $s=6$ and $%
s=4$ at large and small distances between the dipole and the surface of
sphere, respectively.

(ii)~The case $\theta_{p}=\pi/2$. In this case there are two components of
the surface currents which are proportional to $H_{\theta}$ and $H_{\phi}$
at $r=R$. The magnetic pressure is then given by

\begin{equation}
P_{\bot}\left( \theta,\phi\right) =\left. \frac{H_{\theta}^{2}+H_{\phi}^{2}}{%
8\pi}\right\vert _{r=R}=\frac{p^{2}}{8\pi r_{0}^{6}}\frac{\Upsilon
_{1}^{2}\left( \xi,\theta\right) \cos^{2}\phi+\Upsilon_{2}^{2}\left(
\xi,\theta\right) \sin^{2}\phi}{\Upsilon^{6}\left( \xi,\theta\right) }, 
\label{16}
\end{equation}
where

\begin{equation}
\Upsilon_{1}\left( \xi,\theta\right) =\Upsilon_{2}\left( \xi,\theta\right)
\cos\theta-\xi\sin^{2}\theta\left[ \frac{6}{\Upsilon^{2}}-\frac{1}{1-\xi
\cos\theta+\Upsilon}-\frac{\Upsilon\left( 1+\Upsilon\right) }{\left(
1-\xi\cos\theta+\Upsilon\right) ^{2}}\right] ,   \label{17}
\end{equation}

\begin{equation}
\Upsilon_{2}\left( \xi,\theta\right) =\frac{1-\xi^{2}+2\Upsilon}{1-\xi
\cos\theta+\Upsilon},\quad\Upsilon=\sqrt{1+\xi^{2}-2\xi\cos\theta}. 
\label{18}
\end{equation}
At large distances, $\xi\ll1$, the magnetic pressure is maximum at $\phi
\simeq\frac{\pi}{2}$ and $\frac{3\pi}{2}$ (in equatorial plane), and $%
\theta=0$, $\pi$. At small distances, $1-\xi\ll1$, only the region of sphere
with $\theta\sim1-\xi\sim0$ will be strongly deformed.

The total ponderomotive magnetic force acting on the sphere is calculated as

\begin{equation}
\mathcal{F}_{\bot}=R^{2}\int_{0}^{\pi}\sin\theta
d\theta\int_{0}^{2\pi}P_{\bot}\left( \theta,\phi\right) d\phi=\frac{p^{2}}{%
4r_{0}^{4}}\frac {\xi^{2}\left( 3+8\xi^{2}+\xi^{4}\right) }{\left(
1-\xi^{2}\right) ^{4}}.   \label{19}
\end{equation}
Again as for $\theta_{p}=0$, $\pi$ the force $\mathcal{F}_{\bot}$ behaves as 
$\mathcal{F}_{\bot}\sim l^{-s}$ with $s=6$ and $s=4$ at large and small
distances, respectively. However, comparing equations~(\ref{15}) and (\ref%
{19}) we conclude that the total magnetic force at $\theta_{p}=\pi/2$ is
smaller than for parallel or antiparallel orientation of the dipole. For
instance, from equations~(\ref{15}) and (\ref{19}) we obtain $\mathcal{F}%
_{\parallel}\simeq4\mathcal{F}_{\bot}$ and $\mathcal{F}_{\parallel}\simeq2%
\mathcal{F}_{\bot}$ at $\xi\ll1$ and $\xi\sim1$, respectively.

\section{Electrodynamic treatment}

In this section we consider the moving boundary problem of the plasma sphere
expansion in the vacuum. In this sense unlike the magnetostatic problem
considered above it is convenient here to introduce the vector potential of
the induced and dipole magnetic fields. Consider a spherical region of space
containing a neutral infinitely conducting plasma which has expanded at $t=0$
to its present state from a point source located at the point $\mathbf{r}=0$%
. The external space at the point $\mathbf{r}_{0}$ contains a magnetic
dipole $\mathbf{p}$. The magnetic field of this dipole is given by $\mathbf{H%
}_{0}=\mathbf{\nabla}\times\mathbf{A}_{0}$, where the vector potential $%
\mathbf{A}_{0}$ is

\begin{equation}
\mathbf{A}_{0}=\frac{\mathbf{p}\times\left( \mathbf{r}-\mathbf{r}_{0}\right) 
}{\left\vert \mathbf{r}-\mathbf{r}_{0}\right\vert ^{3}}.   \label{20}
\end{equation}

As the spherical plasma cloud expands it both perturbs the external magnetic
field and generates an electric field. Within the spherical plasma region
there is neither an electric field nor a magnetic field. We shall obtain an
analytic solution of the electromagnetic field configuration.

We consider practically interesting case when the vectors $\mathbf{p}$ and $%
\mathbf{r}_{0}$ are parallel (or antiparallel). The general solution for the
arbitrary orientation of $\mathbf{p}$ will be considered in a separate
paper. Within this geometry the problem is symmetric with respect to the
axis $\mathbf{r}_{0}$ which is chosen as the axial axis of the spherical
coordinate system. Then there is only one nonvanishing component of $\mathbf{%
A}_{0}$, $A_{0r}=A_{0\theta}=0$, and

\begin{equation}
A_{0\varphi}=\frac{pr\sin\theta}{\left\vert \mathbf{r}-\mathbf{r}%
_{0}\right\vert ^{3}}=\frac{p}{r_{0}^{2}}\sum_{l=1}^{\infty}D_{l}\left( 
\frac{r}{r_{0}}\right) P_{l}^{1}\left( \cos\theta\right) ,   \label{21}
\end{equation}
where $P_{l}^{\nu}\left( x\right) $ is the generalized Legendre polynomials
with $\nu=1$. Here $D_{l}\left( x\right) =x^{l}$ at $x\leqslant1$ and $%
D_{l}\left( x\right) =x^{-l-1}$ at $x>1$ as defined in Appendix~A.

Since the external region is devoid of free charge density, a suitable gauge
allows the electric and magnetic fields to be derived from the vector
potential $\mathbf{A}$. Having in mind the symmetry of the original dipole
magnetic field it is sufficient to choose the vector potential in the form $%
A_{r}=A_{\theta}=0$,

\begin{equation}
A_{\varphi}\left( r,\theta,t\right) =A_{0\varphi}\left( r,\theta\right)
+\sum_{l=1}^{\infty}\mathcal{A}_{l}\left( r,t\right) P_{l}^{1}\left(
\cos\theta\right) ,   \label{22}
\end{equation}
and the components of the electromagnetic field are given by

\begin{equation}
H_{r}=\frac{1}{r}\frac{\partial A_{\varphi}}{\partial\theta},\quad H_{\theta
}=-\frac{\partial A_{\varphi}}{\partial r},\quad E_{\varphi}=-\frac{1}{c}%
\frac{\partial A_{\varphi}}{\partial t},   \label{23}
\end{equation}
and $H_{\varphi}=E_{r}=E_{\theta}=0$. The equation for $\mathcal{A}%
_{l}\left( r,t\right) $ is obtained from the Maxwell's equations

\begin{equation}
\frac{\partial^{2}\mathcal{A}_{l}}{\partial r^{2}}+\frac{2}{r}\frac {\partial%
\mathcal{A}_{l}}{\partial r}-\frac{l\left( l+1\right) }{r^{2}}\mathcal{A}%
_{l}-\frac{1}{c^{2}}\frac{\partial^{2}\mathcal{A}_{l}}{\partial t^{2}}=0. 
\label{24}
\end{equation}

This equation is to be solved in the external region $r>R\left( t\right) $
subject to the boundary and initial conditions. Here $R\left( t\right) $ is
the plasma sphere radius at the time $t$. The initial conditions are at $t=0$

\begin{equation}
\mathcal{A}_{l}\left( r,0\right) =0,\qquad\frac{\partial\mathcal{A}%
_{l}\left( r,0\right) }{\partial t}=0.   \label{25}
\end{equation}
The first initial condition states that the initial value of $A_{\varphi}$
is that of a dipole magnetic field. The second initial condition states that
there is no initial electric field. Boundary conditions should be imposed at
the spherical surface $r=R\left( t\right) $ and at infinity. Because of the
finite propagation velocity of the perturbed electromagnetic field the
magnetic field at infinity will remain undisturbed for all finite times.
Further, no incoming wave-type solutions are permitted. Thus, for all finite
times $\mathcal{A}_{l}\left( r,t\right) \rightarrow0$ at $r\rightarrow \infty
$. The boundary condition at the expanding spherical surface is $H_{r}=0$
which can be replaced by $A_{\varphi}\left( R\left( t\right)
,\theta,t\right) =0$ or, alternatively,

\begin{equation}
\mathcal{A}_{l}\left( R\left( t\right) ,t\right) =-\frac{p}{r_{0}^{2}}%
D_{l}\left( \frac{R\left( t\right) }{r_{0}}\right) .   \label{26}
\end{equation}

The problem of solving equation~(\ref{24}) subject to the initial and
boundary conditions will be accomplished by the Laplace transform theory.
The Laplace transform $\widetilde{\mathcal{A}}_{l}\left( r,\lambda\right) $
of the function $\mathcal{A}_{l}\left( r,t\right) $ is introduced by

\begin{equation}
\widetilde{\mathcal{A}}_{l}\left( r,\lambda\right) =\int_{0}^{\infty }%
\mathcal{A}_{l}\left( r,t\right) e^{-\lambda t}dt   \label{27}
\end{equation}
with $\mathrm{Re}\lambda>0$. An inverse transformation is established by

\begin{equation}
\mathcal{A}_{l}\left( r,t\right) =\frac{1}{2\pi i}\int_{\sigma-i\infty
}^{\sigma+i\infty}\widetilde{\mathcal{A}}_{l}\left( r,\lambda\right)
e^{\lambda t}d\lambda.   \label{28}
\end{equation}
The real parameter $\sigma$ should be larger than $\mathrm{Re}\lambda_{i}$, $%
\sigma>\mathrm{Re}\lambda_{i}$, where $\lambda_{i}$ are the poles of $%
\widetilde{\mathcal{A}}_{l}\left( r,\lambda\right) $.

The differential equation for $\widetilde{\mathcal{A}}_{l}\left(
r,\lambda\right) $ is found from equations~(\ref{24}) and (\ref{28}) and the
initial conditions in (\ref{25}):

\begin{equation}
\frac{\partial^{2}\widetilde{\mathcal{A}}_{l}\left( r,\lambda\right) }{%
\partial r^{2}}+\frac{2}{r}\frac{\partial\widetilde{\mathcal{A}}_{l}\left(
r,\lambda\right) }{\partial r}-\left[ \frac{l\left( l+1\right) }{r^{2}}+%
\frac{\lambda^{2}}{c^{2}}\right] \widetilde{\mathcal{A}}_{l}\left(
r,\lambda\right) =0.   \label{29}
\end{equation}
Its solution may be written as

\begin{equation}
\widetilde{\mathcal{A}}_{l}\left( r,\lambda\right) =\frac{p}{r_{0}^{2}}\left[
a_{l}\left( \lambda\right) h_{l}^{\left( 1\right) }\left( i\frac{\lambda}{c}%
r\right) +c_{l}\left( \lambda\right) h_{l}^{\left( 2\right) }\left( i\frac{%
\lambda}{c}r\right) \right] ,   \label{30}
\end{equation}
where $h_{l}^{\left( 1\right) }\left( z\right) $ and $h_{l}^{\left( 2\right)
}\left( z\right) $ are the Hankel spherical functions and $a_{l}\left(
\lambda\right) $, $c_{l}\left( \lambda\right) $ are arbitrary functions of $%
\lambda$ determined from the boundary conditions. Since $h_{l}^{\left(
2\right) }\left( z\right) $ gives rise to incoming waves, we should set $%
c_{l}\left( \lambda\right) =0$. The solution to equation~(\ref{24}) at $%
r>R\left( t\right) $ now may be written in the form

\begin{equation}
A_{\varphi}\left( r,\theta,t\right) =\frac{p}{r_{0}^{2}}\sum_{l=1}^{\infty
}P_{l}^{1}\left( \cos\theta\right) \left[ D_{l}\left( \frac{r}{r_{0}}\right)
+\frac{1}{2\pi}\int_{i\sigma-\infty}^{i\sigma+\infty}b_{l}\left(
\lambda\right) h_{l}^{\left( 1\right) }\left( \frac{\lambda}{c}r\right)
e^{-i\lambda t}d\lambda\right] ,   \label{31}
\end{equation}
where $b_{l}\left( \lambda\right) =a_{l}\left( -i\lambda\right) $.

The moving boundary condition in equation~(\ref{26}) requires the
satisfaction of

\begin{equation}
\frac{1}{2\pi i}\int_{i\sigma-\infty}^{i\sigma+\infty}b_{l}\left(
\lambda\right) h_{l}^{\left( 1\right) }\left( \frac{\lambda}{c}R\left(
t\right) \right) e^{-i\lambda t}d\lambda=iD_{l}\left( \frac{R\left( t\right) 
}{r_{0}}\right) .   \label{32}
\end{equation}

Since the sphere moves with a radial velocity $v$ less than the velocity of
light $c$, we have $R<ct$ or $t-R\left( t\right) /c>0$. Thus, the contour in
the integral of equation~(\ref{32}) should be closed by an infinite
semicircle in the lower half plane and the integral evaluated by the method
of residues.

Explicit evaluation of this integral equation~(\ref{32}), may be
accomplished in the special case of a uniform expansion. Choosing the simple
model of constant radial velocity $R\left( t\right) =vt$ and assuming that $%
R\left( t\right) <r_{0}$ equation~(\ref{32}) yields (see Appendix~B for
details)

\begin{equation}
b_{l}\left( \lambda\right) =\frac{\left( -1\right) ^{l}\left( v/r_{0}\right)
^{l}}{\lambda^{l+1}}\frac{i\beta}{\left( 1-\beta^{2}\right) ^{\frac{l+1}{2}}}%
\frac{1}{P_{l}^{-l-1}\left( 1/\beta\right) },   \label{33}
\end{equation}
where $\beta=v/c<1$. Here $P_{\mu}^{\nu}\left( z\right) $ are the
generalized Legendre functions with $z>1$, $\mu=l$, and $\nu=-l-1$.

The solution of equations~(\ref{24}) and (\ref{31}) may be obtained by
inserting equation~(\ref{33}) into (\ref{31}) and evaluating the integral
(see Appendix~B for details). The complete solution may finally be written
in the form at $vt<r<ct$

\begin{equation}
A_{\varphi}\left( r,\theta,t\right) =A_{0\varphi}\left( r,\theta\right) -%
\frac{p}{r_{0}^{2}}\sum_{l=1}^{\infty}\left( \frac{r}{r_{0}}\right) ^{l}%
\frac{p_{l}\left( 1/\zeta\right) }{p_{l}\left( 1/\beta\right) }%
P_{l}^{1}\left( \cos\theta\right) ,   \label{34}
\end{equation}
$A_{\varphi}\left( r,\theta,t\right) =A_{0\varphi}\left( r,\theta\right) $
at $r\geqslant ct$ and $A_{\varphi}\left( r,\theta,t\right) =0$ at $%
r\leqslant vt$. Here $\zeta=r/ct<1$, and

\begin{equation}
p_{l}\left( z\right) =2^{l}l!\left( z^{2}-1\right) ^{\frac{l+1}{2}%
}P_{l}^{-l-1}\left( z\right) =\int_{1}^{z}\left( \tau^{2}-1\right)
^{l}d\tau.   \label{34a}
\end{equation}
The electromagnetic field components can be evaluated according to equation~(%
\ref{23}). From equations~(\ref{23}) and (\ref{34}) it can be easily checked
that the boundary condition on the moving surface, $\mathbf{E}\left(
R\right) =-\frac{1}{c}\left[ \mathbf{v}\times\mathbf{H}\left( R\right) %
\right] $ (or $E_{\varphi}\left( R\right) =-\beta H_{\theta}\left( R\right) $%
), is satisfied automatically. It may also be noted that this special case
of the uniform expansion falls within the conical flow techniques, as
indicated in \cite{kat61} for the case of uniform magnetic field. From
symmetry considerations one seeks a solution of the form $\mathcal{A}%
_{l}\left( r,t\right) =r^{\nu}\Phi\left( r/ct\right) $. Substitution into
the differential equation~(\ref{24}) yields an explicitly solvable ordinary
differential equation whose solution, upon application of the boundary
conditions ($\Phi\left( 1\right) =0$, $\Phi\left( \beta\right)
=-p/r_{0}^{l+2}$), is given by equation~(\ref{34}).

It should be noted that all above results are valid only for $R(t)<r_{0}$ or 
$t<r_{0}/v$. At the time $t=r_{0}/v$ the dipole will enter into the plasma
sphere and hence will be completely shielded by the latter. Therefore at $%
t\geq r_{0}/v$ the total electromagnetic field vanishes and the radiation is
interrupted.

\section{Energy balance}

Previously significant attention has been paid \cite{dit00,rai63} to the
question of what fraction of energy is emitted and lost in the form of
electromagnetic pulse propagating outward of the expanding plasma. In this
section we consider the energy balance during the plasma sphere expansion in
the presence of the magnetic dipole. When the plasma sphere of the zero
initial radius is created at $t=0$ and starts expanding, external magnetic
field $\mathbf{H}_{0}$ is perturbed by the electromagnetic pulse, $\mathbf{H}%
^{\prime}\left( \mathbf{r},t\right) =\mathbf{H}\left( \mathbf{r},t\right) -%
\mathbf{H}_{0}\left( \mathbf{r}\right) $, $\mathbf{E}\left( \mathbf{r}%
,t\right) $, propagating outward with the speed of light. The tail of this
pulse coincides with the moving plasma boundary $r=R\left( t\right) $ while
the leading edge is at $r=ct$. Ahead of the leading edge, the magnetic field
is not perturbed and equals $\mathbf{H}_{0}\left( \mathbf{r}\right) $ while
the electric field is zero.

Our starting point is the energy balance equation (Poynting equation)

\begin{equation}
\mathbf{\nabla}\cdot\mathbf{S}=-\mathbf{j}\cdot\mathbf{E}-\frac{\partial }{%
\partial t}\frac{E^{2}+H^{2}}{8\pi},   \label{35}
\end{equation}
where $\mathbf{S}=\frac{c}{4\pi}\left[ \mathbf{E}\times\mathbf{H}\right] $
is the Poynting vector and $\mathbf{j}=j_{\varphi}\mathbf{e}_{\varphi}$
(with $\left\vert \mathbf{e}_{\varphi}\right\vert =1$) is the azimuthal
surface current density. The energy radiated to infinity is measured as a
Poynting vector integrated over time and over the surface $S_{c}$ of the
sphere with radius $r_{c}<r_{0}$ (control sphere) and the volume $\Omega_{c} 
$ enclosing the plasma sphere ($r_{c}>R$ or $0\leqslant t<r_{c}/v$).
Integrating over time and over the volume $\Omega_{c}$ equation~(\ref{35})
can be represented as

\begin{equation}
W_{S}\left( t\right) =W_{J}\left( t\right) +\Delta W_{\mathrm{EM}}\left(
t\right) ,   \label{36}
\end{equation}
where

\begin{equation}
W_{S}\left( t\right) =2\pi r_{c}^{2}\int_{0}^{t}dt^{\prime}\int_{0}^{\pi
}S_{r}\sin\theta d\theta,\quad W_{J}\left( t\right)
=-\int_{0}^{t}dt^{\prime}\int_{\Omega_{c}}\mathbf{j}\cdot\mathbf{E}d\mathbf{r%
}.   \label{37}
\end{equation}
Here $S_{r}=-\frac{c}{4\pi}E_{\varphi}H_{\theta}$ is the radial component of
the Poynting vector. $W_{\mathrm{EM}}\left( t\right) $ and $\Delta W_{%
\mathrm{EM}}\left( t\right) =W_{\mathrm{EM}}\left( 0\right) -W_{\mathrm{EM}%
}\left( t\right) $ are the total electromagnetic energy and its change (with
minus sign) in a volume $\Omega_{c}$, respectively. $W_{J}\left( t\right) $
is the energy transferred from plasma sphere to electromagnetic field and is
the mechanical work with minus sign performed by the plasma on the external
electromagnetic pressure. At $t=0$ the electromagnetic fields are given by $%
\mathbf{H}\left( \mathbf{r},t\right) =\mathbf{H}_{0}\left( \mathbf{r}\right) 
$ and $\mathbf{E}\left( \mathbf{r},t\right) =0$. Hence $W_{\mathrm{EM}%
}\left( 0\right) $ is the energy of the dipole magnetic field in a volume $%
\Omega_{c}$ and can be calculated from equation~(\ref{3}) by replacing $R$
by $r_{c}$ and setting $\sin\theta_{p}=0$,

\begin{equation}
W_{\mathrm{EM}}\left( 0\right) =\int_{\Omega_{c}}\frac{H_{0}^{2}\left( 
\mathbf{r}\right) }{8\pi}d\mathbf{r}=Q\left( u\right) =\frac{p^{2}}{%
16r_{0}^{3}}\left[ \frac{u\left( 1-u^{4}+8u^{2}\right) }{\left(
1-u^{2}\right) ^{3}}-\frac{1}{2}\ln\frac{1+u}{1-u}\right] ,   \label{38}
\end{equation}
where $u=\frac{r_{c}}{r_{0}}<1$. Then the change of the electromagnetic
energy $\Delta W_{\mathrm{EM}}\left( t\right) $ in a volume $\Omega_{c}$ can
be evaluated as

\begin{equation}
\Delta W_{\mathrm{EM}}\left( t\right) =-\int_{\Omega_{c}}\frac{%
E^{2}+H^{2}-H_{0}^{2}}{8\pi}d\mathbf{r}=Q\left( u\right) -\int_{\Omega
_{c}^{\prime}}\frac{E^{2}+H^{2}}{8\pi}d\mathbf{r}.   \label{39}
\end{equation}
In equation~(\ref{39}) $\Omega_{c}^{\prime}$ is the volume of the control
sphere excluding the volume of the plasma sphere (we take into account that $%
\mathbf{H}\left( \mathbf{r},t\right) =\mathbf{E}\left( \mathbf{r},t\right) =0
$ in a plasma sphere). Hence the total energy flux, $W_{S}\left( t\right) $
given by equation~(\ref{37}) is calculated as a sum of the energy loss by
plasma due to the external electromagnetic pressure and the decrease of the
electromagnetic energy in a control volume $\Omega_{c}$. For
non-relativistic ($\beta\ll1$) expansion of a one-dimensional plasma slab
and for uniform external magnetic field ($\mathbf{H}_{0}=\mathrm{const}$) $%
W_{S}\simeq2W_{J}\simeq2\Delta W_{\mathrm{EM}}$, i.e., approximately the
half of the outgoing energy is gained from the plasma, while the other half
is gained from the magnetic energy \cite{dit00}. In the case of
non-relativistic expansion of highly-conducting spherical plasma in the
uniform magnetic field the outgoing energy $W_{S}$ is distributed between $%
W_{J}$ and $\Delta W_{\mathrm{EM}}$ according to $W_{J}=1.5Q_{0}$ and $%
\Delta W_{\mathrm{EM}}=0.5Q_{0}$ with $W_{S}=2Q_{0}$, where $%
Q_{0}=H_{0}^{2}R^{3}/6$ is the magnetic energy escaped from the plasma
volume \cite{rai63}. Therefore in this case the released electromagnetic
energy is mainly gained from the plasma.

Consider now each energy component $W_{S}\left( t\right) $, $W_{J}\left(
t\right) $ and $\Delta W_{\mathrm{EM}}\left( t\right) $ separately. $%
W_{S}\left( t\right) $ is calculated from equation~(\ref{37}). In the first
expression of equation~(\ref{37}) the $t^{\prime}$-integral must be
performed at $\frac{r_{c}}{c}\leqslant t^{\prime}\leqslant t$ ($t<\frac{r_{c}%
}{v}$) since at $0\leqslant t^{\prime}<\frac{r_{c}}{c}$ the electromagnetic
pulse does not reach to the control surface yet and $S_{r}\left(
r_{c}\right) =0$. From equations~(\ref{23}), (\ref{34}) and (\ref{37}) we
obtain

\begin{equation}
W_{S}\left( t\right) =Q\left( u\right) +\frac{p^{2}}{2r_{0}^{3}}\sum
_{l=1}^{\infty}\frac{l\left( l+1\right) }{2l+1}u^{2l+1}\left\{ \frac{\left(
1/\eta^{2}-1\right) ^{2l+1}}{\left( 2l+1\right) p_{l}^{2}\left(
1/\beta\right) }-\left( l+1\right) \left[ \frac{p_{l}\left( 1/\eta\right) }{%
p_{l}\left( 1/\beta\right) }-1\right] ^{2}\right\} ,   \label{40}
\end{equation}
where $\eta=r_{c}/ct<1$. In non-relativistic limit, $\beta\rightarrow0$,
using the asymptotic expression (see, e.g., \cite{gra80}) $p_{l}\left(
z\right) =z^{2l+1}/\left( 2l+1\right) $ at $z\rightarrow\infty$, from
equation~(\ref{40}) we obtain

\begin{align}
W_{S}\left( t\right) & =2Q\left( \xi\right) -Q\left( \kappa\right) +\frac{%
p^{2}}{r_{0}^{3}}\frac{\kappa^{3}}{\left( 1-\kappa^{2}\right) ^{3}}
\label{42} \\
& =\frac{p^{2}}{16r_{0}^{3}}\left[ \frac{2\xi\left( 1+8\xi^{2}-\xi
^{4}\right) }{\left( 1-\xi^{2}\right) ^{3}}+\frac{\kappa\left( \kappa
^{4}+8\kappa^{2}-1\right) }{\left( 1-\kappa^{2}\right) ^{3}}-\frac{1}{2}\ln%
\frac{\left( 1-\kappa\right) \left( 1+\xi\right) ^{2}}{\left(
1+\kappa\right) \left( 1-\xi\right) ^{2}}\right]  \notag
\end{align}
with $\kappa=R^{2}/r_{0}r_{c}$. In equation~(\ref{42}) $Q\left(
\kappa\right) $ represents the magnetic energy of the dipole field in a
sphere having the radius $R_{\ast}=R^{2}/r_{c}<R$ and enclosed in the plasma
sphere.

Next, we calculate the energy loss $W_{J}\left( t\right) $ by the plasma
which is determined by the surface current density, $\mathbf{j}$. From the
symmetry reason it is clear that this current has only azimuthal component
and is localized within thin spherical skin layer, $R-\delta<r<R+\delta$
with $\delta\rightarrow0$, near plasma boundary. Therefore in equation~(\ref%
{37}) the volume $\Omega_{c}$ can be replaced by the volume $%
\Omega_{\delta}\sim R^{2}\delta$ which includes the space between the
spheres with $r=R-\delta$ and $r=R+\delta$. The surface current density is
calculated from the Maxwell's equation, $\mathbf{j}=\left( 1/4\pi\right)
\left( c\mathbf{\nabla}\times\mathbf{H}-\frac{\partial\mathbf{E}}{\partial t}%
\right) $. Within the skin layer we take into account that $\mathbf{E}=-%
\frac{1}{c}\left[ \mathbf{v}\times\mathbf{H}\right] $ and $H_{r}\left(
R\right) =0$. Then

\begin{align}
Q_{J}\left( t\right) & =-\int_{\Omega_{\delta}}\mathbf{j}\cdot \mathbf{E}d%
\mathbf{r}=\frac{1}{4\pi}\int_{\Omega_{\delta}}\mathbf{v\cdot }\left[ 
\mathbf{H}\times\left( \mathbf{\nabla}\times\mathbf{H}\right) \right] d%
\mathbf{r}+\frac{1}{8\pi}\int_{\Omega_{\delta}}\frac{\partial \mathbf{E}^{2}%
}{\partial t}d\mathbf{r}  \label{43} \\
& =v\int\nolimits_{S_{R}}\frac{H_{\theta}^{2}\left( R\right) -E_{\varphi
}^{2}\left( R\right) }{8\pi}dS=\frac{v}{\gamma^{2}}\int\nolimits_{S_{R}}%
\frac{H_{\theta}^{2}\left( R\right) }{8\pi}dS,  \notag
\end{align}
where $\gamma^{-2}=1-\beta^{2}$ and $S_{R}$ are the relativistic factor and
the surface of the expanding plasma, respectively. Note that the moving
boundary modifies the surface current which is now proportional to $%
\gamma^{-2}$ \cite{jac75}. In equation~(\ref{43}) the term with $\frac{%
\partial\mathbf{E}^{2}\left( \mathbf{r},t\right) }{\partial t}$ has been
transformed to the surface integral using the fact that the boundary of the
volume $\Omega_{\delta}$ moves with a constant velocity $v$ and the
electrical field has a jump across the plasma surface. Equation~(\ref{43})
shows that the energy loss by the plasma per unit time is equal to the work
performed by the plasma on the external electromagnetic pressure. This
external pressure is formed by the difference between magnetic and electric
pressures, i.e., the induced electric field tends to decrease the force
acting on the expanding plasma surface. The total energy loss by the plasma
sphere is calculated as

\begin{equation}
W_{J}\left( t\right) =\int_{0}^{t}Q_{J}\left( t^{\prime}\right) dt^{\prime}=%
\frac{p^{2}}{2r_{0}^{3}}\sum_{l=1}^{\infty}\frac{l\left( l+1\right) }{\left(
2l+1\right) ^{2}}\left( \frac{\xi}{\beta^{2}\gamma ^{2}}\right) ^{2l+1}\frac{%
1}{p_{l}^{2}\left( 1/\beta\right) },   \label{44}
\end{equation}
where $\xi=R/r_{0}$. In a non-relativistic case equation~(\ref{44}) yields:

\begin{equation}
W_{J}\left( t\right) =\frac{p^{2}}{r_{0}^{3}}\frac{\xi^{3}}{\left(
1-\xi^{2}\right) ^{3}}.   \label{45}
\end{equation}

\begin{figure}[tbp]
\includegraphics[width=14cm]{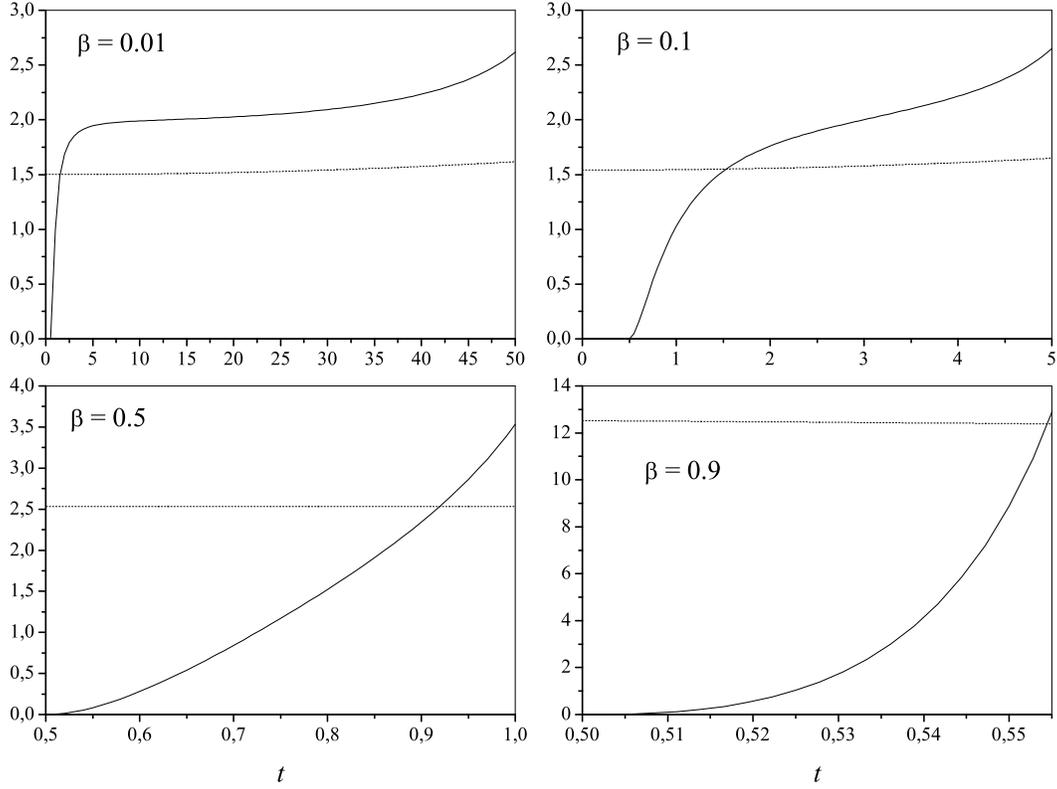}
\caption{The ratios $\Gamma _{S}\left( t\right) $ (solid lines) and $\Gamma
_{J}\left( t\right) $ (dashed lines) for four values of $\protect\beta $ as
a function of $t$ (in units of $r_{0}/c$) calculated from expressions (%
\protect\ref{40}) and (\protect\ref{44}) with $r_{c}=0.5r_{0}$.}
\end{figure}

The change of the electromagnetic energy in a control sphere is calculated
from equation~(\ref{39}). At $R<r_{c}<ct$ (the electromagnetic pulse fills
the whole control sphere) we obtain

\begin{align}
\Delta W_{\mathrm{EM}}\left( t\right) & =Q\left( u\right) -\frac{p^{2}}{%
2r_{0}^{3}}\sum_{l=1}^{\infty}\frac{l\left( l+1\right) }{\left( 2l+1\right)
^{2}}\left( \frac{\xi}{\beta^{2}\gamma^{2}}\right) ^{2l+1}\frac{1}{%
p_{l}^{2}\left( 1/\beta\right) }  \label{46} \\
& +\frac{p^{2}}{2r_{0}^{3}}\sum_{l=1}^{\infty}\frac{l\left( l+1\right) }{2l+1%
}u^{2l+1}\left\{ \frac{\left( 1/\eta^{2}-1\right) ^{2l+1}}{\left(
2l+1\right) p_{l}^{2}\left( 1/\beta\right) }-\left( l+1\right) \left[ \frac{%
p_{l}\left( 1/\eta\right) }{p_{l}\left( 1/\beta\right) }-1\right]
^{2}\right\} .  \notag
\end{align}
Comparing equations~(\ref{40}), (\ref{44}) and (\ref{46}) we conclude that $%
\Delta W_{\mathrm{EM}}\left( t\right) +W_{J}\left( t\right) =W_{S}\left(
t\right) $ as predicted by the energy balance equation~(\ref{36}). The
non-relativistic limit of equation~(\ref{46}) can be evaluated from
equations~(\ref{42}) and (\ref{45}) using the relation $\Delta W_{\mathrm{EM}%
}\left( t\right) =W_{S}\left( t\right) -W_{J}\left( t\right) $. As an
example in figure~1 we show the results of model calculations for the ratios 
$\Gamma_{S}\left( t\right) =W_{S}\left( t\right) /Q_{0}\left( t\right) $ and 
$\Gamma_{J}\left( t\right) =W_{J}\left( t\right) /Q_{0}\left( t\right) $ as
a function of time ($r_{c}/c\leqslant t<r_{c}/v$). Here $Q_{0}\left(
t\right) =Q\left( \xi\right) $ is the dipole magnetic energy escaped from
the plasma sphere. For the relativistic factor $\beta$ we have chosen a wide
range of values. We recall that at $0\leqslant t\leqslant r_{c}/c$, i.e. the
electromagnetic pulse does not yet reach to the surface of the control
sphere, $W_{S}\left( t\right) =0$. Unlike the case with uniform magnetic
field discussed above (see also \cite{dit00,rai63}) there are no simple
relations between the energy components $W_{S}\left( t\right) $, $%
W_{J}\left( t\right) $ and $Q_{0}\left( t\right) $. However, at the initial
stage ($t\ll r_{c}/v$) of non-relativistic expansion the dipole field at
large distances can be treated as uniform and the energies $W_{S}\left(
t\right) $ and $W_{J}\left( t\right) $ are close to the values $2Q_{0}\left(
t\right) $ and $1.5Q_{0}\left( t\right) $ (see figure~1), respectively. For
any $\beta$ the ratio $\Gamma_{J}\left( t\right) $ is almost constant and
may be approximated as $\Gamma_{J}\left( t\right) \simeq\Gamma_{J}\left(
0\right) $ or alternatively $W_{J}\left( t\right) \simeq1.5CQ_{0}\left(
t\right) $, where $C=\gamma^{-6}\left( 1-\beta \right) ^{-4}\left(
1+2\beta\right) ^{-2}$ is some kinematic factor. For $\beta\sim1$ this
factor is very large and behaves as $C\simeq\left( 8/9\right) \left(
1-\beta\right) ^{-1}\gg1$. As expected the total energy flux, $W_{S}\left(
t\right) $, increases monotonically with $t$. At the final stage ($t=r_{c}/v$%
) of relativistic expansion (with $\beta\sim1$) $W_{S}\simeq W_{J}$. Hence
in this case the radiated energy $W_{S}$ is mainly gained from the plasma
sphere.

\section{Conclusion}

An exact solution of the uniform radial expansion of a neutral, infinitely
conducting plasma sphere in the presence of a dipole magnetic field has been
obtained. The electromagnetic fields are derived by using the appropriate
boundary and initial conditions, equations~(\ref{25}) and (\ref{26}). It is
shown that the electromagnetic fields are perturbed only within the domain
extending from the surface of the expanding plasma sphere $r=R=vt$ to the
surface of the expanding information sphere $r=ct$. External to the sphere $%
r=ct$ the magnetic field is not perturbed and is given by the dipole
magnetic field. In the course of this study we have also considered the
energy balance during the plasma sphere expansion. The model calculations
show that the radiated energy is mainly gained from the plasma sphere. For
relativistic expansion the ratio $W_{S}/W_{J}$ is close to unity and the
radiated energy is practically gained only from plasma sphere.

We expect our theoretical findings to be useful in experimental
investigations as well as in numerical simulations of the plasma expansion
into ambient nonuniform magnetic field. One of the improvements of our model
will be to include the effect of the deceleration of the plasma sphere as
well as the derivation of the dynamical equation for the surface
deformation. A study of this and other aspects will be reported elsewhere.

\begin{acknowledgments}
This work has been partially supported by the Armenian Ministry of Higher
Education and Science Grant.
\end{acknowledgments}

\appendix

\section{Sums with Legendre polynomials}

Using the known relation \cite{gra80}

\begin{equation}
F_{0}\left( x,\theta\right) =\frac{1}{\left( 1+x^{2}-2x\cos\theta\right)
^{1/2}}=\sum_{l=0}^{\infty}D_{l}\left( x\right) P_{l}\left( \cos
\theta\right) ,   \label{A1}
\end{equation}
where $D_{l}\left( x\right) =x^{l}$ at $\left\vert x\right\vert \leqslant1$
and $D_{l}\left( x\right) =x^{-l-1}$ at $\left\vert x\right\vert >1$ one can
derive some sums with Legendre polynomials $P_{l}\left( \cos\theta\right) $
which are used in the main text of the paper. The first relation is obtained
from equation~(\ref{A1}) by taking the partial derivative of the function $%
F_{0}\left( x,\theta\right) $:

\begin{equation}
\frac{\partial}{\partial x}F_{0}\left( x,\theta\right) =\frac{\cos\theta -x}{%
\left( 1+x^{2}-2x\cos\theta\right) ^{3/2}}=\sum_{l=0}^{\infty}D_{l}^{\prime}%
\left( x\right) P_{l}\left( \cos\theta\right) .   \label{A2}
\end{equation}
Here the prime indicates the derivative with respect to the argument.

The second relation follows from equation~(\ref{A1}) if we take the partial
derivative over $\theta$:

\begin{equation}
-\frac{\partial}{\partial\theta}F_{0}\left( x,\theta\right) =\frac {%
x\sin\theta}{\left( 1+x^{2}-2x\cos\theta\right) ^{3/2}}=\sum_{l=1}^{\infty
}D_{l}\left( x\right) P_{l}^{1}\left( \cos\theta\right) ,   \label{A3}
\end{equation}
where $P_{l}^{1}\left( \cos\theta\right) $ are the generalized Legendre
polynomials $P_{l}^{\nu}\left( \cos\theta\right) $ with $\nu=1$.

The third sum is calculated as

\begin{equation}
\frac{\partial}{\partial x}\left[ xF_{0}\left( x,\theta\right) \right] =%
\frac{1-x\cos\theta}{\left( 1+x^{2}-2x\cos\theta\right) ^{3/2}}=\sum
_{l=0}^{\infty}\left[ xD_{l}^{\prime}\left( x\right) +D_{l}\left( x\right) %
\right] P_{l}\left( \cos\theta\right) ,   \label{A4}
\end{equation}

Consider now the sum

\begin{equation}
F\left( x,\theta\right) =\sum_{l=1}^{\infty}\frac{l}{l+1}x^{l+1}P_{l}^{1}%
\left( \cos\theta\right) =-\frac{\partial}{\partial\theta}\sum
_{l=1}^{\infty}\frac{l}{l+1}x^{l+1}P_{l}\left( \cos\theta\right) , 
\label{A5}
\end{equation}
where $x<1$. It is easy to see that

\begin{equation}
\frac{\partial}{\partial x}F\left( x,\theta\right) =-x\frac{\partial^{2}}{%
\partial x\partial\theta}F_{0}\left( x,\theta\right) =x\sin\theta \frac{%
\partial}{\partial x}\frac{x}{\left( 1+x^{2}-2x\cos\theta\right) ^{3/2}}. 
\label{A6}
\end{equation}
Using equation~(\ref{A6}) we finally obtain

\begin{align}
F\left( x,\theta\right) & =\sin\theta\left[ x^{2}F_{0}^{3}\left(
x,\theta\right) -\int_{0}^{x}F_{0}^{3}\left( t,\theta\right) tdt\right] 
\notag \\
& =\frac{x^{2}\sin\theta}{\left( 1+x^{2}-2x\cos\theta\right) ^{3/2}}-\frac{1%
}{\sin\theta}\left( 1-\frac{1-x\cos\theta}{\left(
1+x^{2}-2x\cos\theta\right) ^{1/2}}\right) .   \label{A7}
\end{align}
In equation~(\ref{A7}) we have used the initial condition $F\left(
0,\theta\right) =0$.

\section{Evaluation of the vector potential}

For evaluation of the integral equation~(\ref{32}) we consider the explicit
expression for the spherical Hankel functions $h_{l}^{\left( 1\right)
}\left( z\right) $ \cite{gra80}

\begin{equation}
h_{l}^{\left( 1\right) }\left( z\right) =\left( -i\right)
^{l+1}e^{iz}\sum_{k=0}^{l}\left( \frac{i}{2}\right) ^{k}\frac{\left(
l+k\right) !}{k!\left( l-k\right) !}\frac{1}{z^{k+1}}   \label{B1}
\end{equation}
and assume that $b_{l}\left( \lambda\right) =B_{l}/\lambda^{l+1}$, where $%
B_{l}$ does not depend on $\lambda$. This choice of $b_{l}\left(
\lambda\right) $ assures that $B_{l}$ is constant (see below). Inserting
equation~(\ref{B1}) and $b_{l}\left( \lambda\right) $ into equation~(\ref{32}%
) we obtain

\begin{equation}
B_{l}\sum_{k=0}^{l}\left( \frac{i}{2}\right) ^{k}\frac{\left( l+k\right) !}{%
k!\left( l-k\right) !}\left( \frac{1-\beta}{\beta\tau}\right)
^{k+1}\Im_{k+l+1}\left( \tau\right) =-i^{l}\left( \frac{vt}{r_{0}}\right)
^{l},   \label{B2}
\end{equation}
where $\tau=t\left( 1-\beta\right) >0$ and

\begin{equation}
\Im_{n}\left( \tau\right) =\frac{1}{2\pi i}\int_{i\sigma-\infty}^{i\sigma+%
\infty}\frac{e^{-i\lambda\tau}d\lambda}{\lambda^{n+1}}=\frac{1}{n!}\frac{%
\partial^{n}}{\partial q^{n}}\left[ \frac{1}{2\pi i}\int
_{i\sigma-\infty}^{i\sigma+\infty}\frac{e^{-i\lambda\tau}d\lambda}{\lambda -q%
}\right] _{q=0}.   \label{B3}
\end{equation}
Here $\mathrm{Im}q<\sigma$. The integral within the square brackets
according to the Kochi's theorem and at $\tau>0$ is equal to $-e^{-iq\tau}$.
Therefore $\Im_{n}\left( \tau\right) =-\left( -i\tau\right) ^{n}/n!$.
Inserting this function into equation~(\ref{B2}) we arrive at equation~(\ref%
{33}) (see, e.g., \cite{gra80}). The complete solution is obtained by
inserting equation~(\ref{33}) into equation~(\ref{31}) and evaluating the
contour integral as it was done above.

\end{document}